# Parity-violating macroscopic force between chiral molecules and source mass

Yonghong Hu，Zhongzhu Liu，Qing Xu，Jun Luo[*]

Department of Physics, Huazhong University of Science and Technology, Wuhan 430074, People's Republic of China

A theory concerning non-zero macroscopic chirality-dependent force between a source mass and homochiral molecules due to the exchange of light particles is presented in this paper. This force is proposed to have opposite sign for molecules with opposite chirality. Using the central field approximation, we calculate this force between a copper block and a vessel of homochiral molecules (methyl phenyl carbinol nitrite). The magnitude of force is estimated with the published limits of the scalar and pseudo-scalar coupling constants. Based on our theoretical model, this force may violate the equivalence principle when the homochiral molecules are used to be the test masses.

## Ⅰ.INTRODUCTION

The local equivalence of the inertial force and the gravitational force is so-called equivalence principle, which is one of the most fundamental principles in nature. As it is an important cornerstone in physics, people keep concerning every possible effect that may violate it [1]. Recently, the force mediated through exchange of axions or other particles becomes a focus of research concerning the violation of the equivalence principle [2–4]. This proposed force is spin-dependent so that it was designed to be detected in experiments by using magnetized or polarized test masses [3]. However, the spin-dependent force is expected to be very small, one must avoid any electromagnetic effect in experiment in order to test precisely the equivalence principle. Therefore, it is necessary to explore the nature of force mediated through exchanging axions [4] which violates the equivalence principle with non-magnetized test mass.

Chirality is a basic feature of configuration of some material, crystals and organic molecules. These materials and their mirror image enantiomers have the same composition but different geometrical structure, left screw or right screw. Chiral medium has the rotatory power that the polarization plane of a linearly polarized light transmitting through it will be rotated [5]. A parity non-conserving energy difference between the enantiomers of a chiral molecule due to the weak neutral currents had been reported [6].The coupling of an axion to matter is pseudoscalar that its sign is changed by a space reflection. So the coupling of axions to matter with opposite chirality has the opposite sign, which may have some observable effects.

In this paper, we present the potential between an achiral source mass and a vessel of homochiral (left-handed or right-handed) molecules. Acting together with the spin-orbit interaction, such potential may induce a macroscopic force such that it will change its direction when the chirality of the molecules is changed. According to the proposed theory, this force may violate the equivalence principle as the test mass is a vessel of homochiral molecules and the source mass is an achiral material. To demonstrate this effect, we estimate the magnitude of the force between a copper block and a cubical vessel of homochiral molecules using the central field approximation. Its possible influence to gravitational experiments is discussed as well.

---

[*] Electronic address: junluo@mail.hust.edu.cn

## Ⅱ. THEORY AND MODELING
### A. Electron-Nucleon Interaction

In the non-relativity approximation, a potential between an electron and a nucleon will present due to the exchange of much-low-mass pseudo-scalar particles as follows [3]

$$H_{int} = \hbar(g_s g_p)\frac{\sigma \cdot \hat{r}}{8\pi m_e c}(\frac{m_\varphi}{r} + \frac{1}{r^2})e^{-m_\varphi r}, \tag{1}$$

where $c$ is the light speed in vacuum, $\hbar\sigma/2$ is the electron spin, $r$ is the displacement vector between the nucleon and the electron, $g_s$ and $g_p$ are the scalar and pseudo-scalar coupling constants respectively, $m_e$ is the mass of the electron and $m_\varphi$ is the mass of exchanged pseudo-scalar particle. The range of the interaction is given as $\lambda = \hbar/m_\varphi c$.

### B. Chirality Dependent Potential between a Source Mass and Chiral Molecules

Chirality is the fundamental character of wide kinds of molecules and has been studied for a few centuries. Within electron system of a chiral molecule, the distribution of valence electrons displays the chiral configuration of the molecule. So let's study the valence electrons in a vessel of homochiral molecules interacting with the nucleons in an achiral source mass. A macroscopic potential between the nucleons in the source mass and the valence electrons in the homochiral molecules is given by the sum of all the potentials of form (1) between electron-nucleon pairs. It will be zero due to the zero total magnetic moment of the molecules. Considering the spin-orbital interaction of the valence electrons in a quantum calculation, the situation will change that the macroscopic potential will has a non-zero value. And the potential will have the same magnitude but opposite sign if the chirality of the molecules is changed.

Since the valence electrons and the nucleons in our study belong to two macroscopic bodies respectively, we can study the motions of valence electrons in the molecules with the adiabatic approximation, which means the ions of the molecules are regarded as fixed [7]. Then the total Hamiltonian of the valence electron system in the homochiral molecules can be written as

$$\hat{H} = \hat{H}_0 + \hat{H}_1 + \hat{H}_2, \tag{2}$$

where $\hat{H}_1 = \sum_{i=1}^{M}\sum_{j=1}^{N} \frac{g_s g_p \hbar}{8\pi m_e c}(\sigma_i \cdot \hat{r}_{ij})(m_\varphi + \frac{1}{r_{ij}})\frac{e^{-r_{ij}m_\varphi}}{r_{ij}}$ is the interaction Hamiltonian introduced by the potential (1) between all the nucleons in the source mass and valence electrons in the chiral molecules, $\hat{H}_2 = -\sum_{i=1}^{M} B(r_i) \cdot \sigma_i = \sum_{i=1}^{M} \xi(r_i) l_i \cdot \sigma_i$ represents the orbit-spin interaction Hamiltonian of the valence electron system [8] and $\hat{H}_0$ is the Hamiltonian of the valence electron system involving only the Coulomb interaction. $B(r_i)$ is the magnetic induction at the position of the $i$th valence electron, $l_i$ and $\sigma_i/2$ are the orbital angular momentum and the spin of the $i$th valence electron. And $\xi(r) = \frac{\hbar^2}{m_e^2 c^2}\frac{1}{r}\frac{\partial U(r)}{\partial r}$ is the orbit-spin coupling coefficient for electron in a Coulomb potential $U(r)$.

The energy shift of the valence electron system due to the Hamiltonian $\hat{H}_1$ and $\hat{H}_2$ can be given with the stationary state perturbation theory [9]. A simple calculation in the second order correction of the energy shows that the chirality-dependent potential induced by the Hamiltonian $(\hat{H}_1 + \hat{H}_2)$ is [10]

$$E_V = 2\,\text{Re} \sum_{n \neq 0} \frac{\langle \Psi_n | H_1 | \Psi_0 \rangle \langle \Psi_0 | H_2 | \Psi_n \rangle}{E_0 - E_n}. \quad (3)$$

Where $|\Psi\rangle$ is the wave function of valence electron system in the molecules, the subscript "0" represents the initial state and "n" represents finial state. The calculation of the potential (3) can be carried out in the single-electron approximation writing the system wave function $|\Psi\rangle$ as the product of single-electron wave functions [11]. Then, we further factor each single-electron wave function as the product of its orbital wave function $|\psi\rangle$ and the spin wave function $|m\rangle$ [12]. After the summation of spin magnetic quantum number employing the relation of $(\sigma \cdot A)(\sigma \cdot B) = A \cdot B + i\sigma \cdot A \times B$, the expression (3) reduces to the following form,

$$E_V = 2\,\text{Re} \sum_i \sum_j \sum_{n \neq 0} \frac{1}{E_{i0} - E_{in}} \{ \langle \psi_{in} | \frac{g_s g_p \hbar}{8\pi m_e c}(m_\varphi + \frac{1}{r_{ij}})\frac{e^{-r_{ij}m_\varphi}}{r_{ij}^2} r_{ij} | \psi_{i0} \rangle \cdot \langle \psi_{i0} | \xi(r_i) l_i | \psi_{in} \rangle \}. \quad (4)$$

Where $r_{ij}$ is the displacement vector between the $i$th electron and the $j$th nucleon.,

$|\psi_{i0}\rangle$ and $|\psi_{in}\rangle$ are the initial and final states of the ith valence electron and $E_{i0}$ and $E_{in}$ are their eigen-energies respectively. Let $|\psi_L\rangle$ be a state of left-handed molecule. Under the space reflection $P$, it becomes a state of the right-handed molecule $|\psi_R\rangle$ as $|\psi_R\rangle = P|\psi_L\rangle$. The orbital angular momentum $l_i$ is an axial vector and it doesn't change under the reflection. But $r_{ij}$ will change its sign as a polar vector. Therefore the products of matrix elements in left- and right- handed states have the following relation:

$$\langle\psi_{Ln}|H_1|\psi_{L0}\rangle\langle\psi_{L0}|H_2|\psi_{Ln}\rangle = -\langle\psi_{Rn}|H_1|\psi_{R0}\rangle\langle\psi_{R0}|H_2|\psi_{Rn}\rangle \qquad (5)$$

This means the energy $E_V$ takes the opposite value for the molecules with opposite chirality and the same achiral source mass.

### III. CALCULATION AND RESULTS
#### A. Crude Estimation of Matrix Elements

This potential can be estimated quantitatively. We take $(E_0 - E_n) \approx 1$ eV for a rough estimation [13]. We can take the energy split due to the spin-orbit interaction as the value of the matrix element $\langle\psi_0|\frac{\hbar}{2}\sigma\cdot\xi(r)l|\psi_n\rangle$ approximately [14], whose value can be read from published theoretical calculations that $\frac{\hbar}{2}\langle\psi_0|\sigma\cdot\xi(r)l|\psi_n\rangle \approx 0.01$ eV [15]. So, we get the estimation

$$\frac{1}{(E_0 - E_n)}\langle\psi_0|\frac{\hbar}{2}\sigma\cdot\xi(r)l|\psi_n\rangle \approx 10^{-2}. \qquad (6)$$

Suppose the distance between a nucleon and an electron is a half millimeter and the mass of pseudo-scalar exchanged particle is $10^3$. Then we get the potential between such a nucleon and electron as follows

$$\langle\psi_n|\frac{g_s g_p \hbar \sigma \cdot r}{8\pi m_e c}(m_\varphi + \frac{1}{r})\frac{e^{-rm_\varphi}}{r^2}|\psi_0\rangle \approx 2.8\times 10^{-8} g_s g_p . \qquad (7)$$

With the chirality factor $\chi = 10^{-3}$ measuring the extent of asymmetry of the chiral molecule [13], the potential between a nucleon and an electron is about $2.8\times 10^{-13} g_s g_p$. Further we can study the macroscopic force between macroscopic bodies with large amount of nucleons and electrons.

**B. Calculation of the Macroscopic Force**

In the calculation above, the matrix element was estimated crudely. In fact the matrix element can be computed through quantum mechanical methods. However, we care mainly for the effect of molecular asymmetry, so some simplifications can be adopted for some chiral molecules. For molecule having a centre, we can take one simple quantum method, the central field approximation presented in the crystal field theory of the quantum chemistry [7]. These kinds of chiral molecules are made up of central atoms and ligands. And we further look for the ligands and the central atoms as ideal point-like particles. Let the origin of coordinates be on the center of the central atom in the molecule. Let $r, \theta, \varphi$ and $R, \Theta, \Phi$ be the spherical coordinates of valence electron and ligands in the molecule respectively. Then the Coulomb potential of a valence electron is represented as [7]

$$U(r) = \sum_{l=0}^{\infty} \sum_{m=-l}^{l} \frac{4\pi}{2l+1} Y_l^m(\theta,\varphi) \int \rho(R,\Theta,\Phi) \frac{r_<^l}{r_>^{l+1}} [Y_l^m(\Theta,\Phi)]^* d\tau \quad , \tag{8}$$

where $r_>$ and $r_<$ represent the bigger one and the smaller one of $r$ and $R$ respectively. $Y_l^m(\theta,\varphi)$ is the usual spherical harmonics, and $\rho(R,\Theta,\Phi)$ is the charge density of ligands in the molecule.

The wave function of single electron is written as the product of the radial component and the angular component $|\psi\rangle = R_{CL}(r) \cdot Y_{LM}(\theta,\varphi)$. Where $C$, $L$ and $M$ are the main quantum number, the angular moment quantum number and the spin magnetic quantum number respectively. However, functions $R_{CL}(r)$ and $Y_{LM}(\theta,\varphi)$ are difficult to determine accurately. For the sake of simplicity, the single-electron wave function of electron is further replaced by the electron wave function of classical hydrogen-like atom in our practical calculation next. And these matrix elements in the potential (4) can be given according to elementary quantum mechanics and the coupling law of angular momentums.

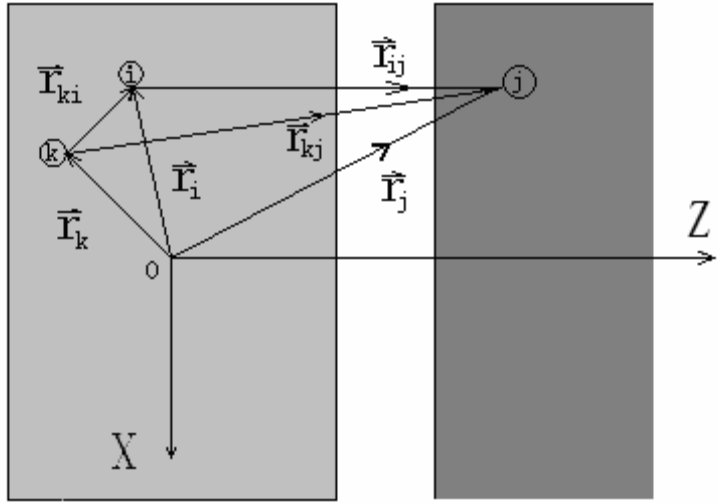

FIG.1. the geometric configuration of the copper cube and the cubical vessel filled with homochiral molecules and the scheme of position vectors of the central atom of the kth molecule, the ith valence electron of the molecule in the vessel and the jth nucleon in the copper cube. The origin of the coordinate is put at the mass center of the vessel.

A copper cube and a cubical vessel filled with homochiral molecules have the same section, but different thickness as shown in Fig.1. Let the copper cube has $N_1$ nucleons and there are $N_2$ chiral molecules in the vessel. Each molecule has $N_3$ valence electrons. The interaction between them can be counted with the central field approximation. The coordinate origin is placed at the center of the vessel and the z axle is along the normal line of the parallel sections. In the coordinate, the kth chiral molecule has the central atom with the position vector $r_k$, its ith valence electron has the position vector $r_i$. The jth nucleon in the copper cube has the position vector $r_j$. The displacement vector from the central atom of the kth chiral molecule to jth nucleon is $r_{kj}$. And the displacement vector of the ith valence electron with respect to the central atom is $r_{ki}$ and we have the equation $r_{ij} = r_{kj} - r_{ki}$. The vector $r_{kj}$ has a macroscopic scale, but the vector $r_{ki}$ has only a microscopic scale, so $|r_{ij}| \gg |r_{ki}|$ and we can take $|r_{ij}| \approx |r_{kj}|$ approximately. The electron wave function and the spin-orbit coupling are both the functions of $r_{ki}$. Therefore, the product of matrix elements in equation (4) can be written as

$$\langle\psi_{in}|(m_\varphi+\frac{1}{r_{ij}})\frac{e^{-r_{ij}m_\varphi}}{r_{ij}^2}\mathbf{r}_{ij}|\psi_{i0}\rangle\cdot\langle\psi_{i0}|\xi(\mathbf{r}_i)\mathbf{l}_i|\psi_{in}\rangle \approx -\langle\psi_{in}|(m_\varphi+\frac{1}{r_{kj}})\frac{e^{-r_{kj}m_\varphi}}{r_{kj}^2}\mathbf{r}_{ki}|\psi_{i0}\rangle\cdot\langle\psi_{i0}|\xi(\mathbf{r}_{ki})\mathbf{l}_i|\psi_{in}\rangle$$

$$= -(m_\varphi+\frac{1}{r_{kj}})\frac{e^{-r_{kj}m_\varphi}}{r_{kj}^2}\langle\psi_{in}|\mathbf{r}_{ki}|\psi_{i0}\rangle\cdot\langle\psi_{i0}|\xi(\mathbf{r}_{ki})\mathbf{l}_i|\psi_{in}\rangle$$

(9)

According to the chirality-dependent potential (4), the macroscopic force between the copper cube and the cubical vessel of homochiral molecules along the z axle can be expressed as

$$F_z = -\frac{g_s g_p \hbar}{4\pi m_e c}\cdot\sum_{i=1}^{N_2\cdot N_3}\sum_{j=1}^{N_1}\frac{(m_\varphi^2 r_{ij}^2+3m_\varphi r_{ij}+3)|z_j-z_i|e^{-r_{ij}m_\varphi}}{r_{ij}^5}\cdot\mathrm{Re}\sum_{n\neq 0}\frac{\langle\psi_{in}|\mathbf{r}_{ij}|\psi_{i0}\rangle\cdot\langle\psi_{i0}|\xi(\mathbf{r}_{ij})\mathbf{l}_i|\psi_{in}\rangle}{E_{i0}-E_{in}}$$

$$\approx -\frac{g_s g_p \hbar}{4\pi m_e c}\cdot\sum_{k=1}^{N_2}\sum_{j=1}^{N_1}\sum_{i=1}^{N_3}\frac{(m_\varphi^2 r_{kj}^2+3m_\varphi r_{kj}+3)|z_j-z_k|e^{-r_{kj}m_\varphi}}{r_{kj}^5}\cdot\mathrm{Re}\sum_{n\neq 0}\frac{\langle\psi_{in}|\mathbf{r}_{ki}|\psi_{i0}\rangle\cdot\langle\psi_{i0}|\xi(\mathbf{r}_{ki})\mathbf{l}_i|\psi_{in}\rangle}{E_{i0}-E_{in}}$$

(10)

The summations over all valence electrons and nucleons can be done through integration over the volume of the copper cube and the cubical vessel under the assumption that both the homochiral molecules and the nucleons distribute uniformly in the vessel and the copper cube.

**C. Macroscopic Force with Concrete Chiral Molecules**

Now we calculate the force for homochiral molecule methyl phenyl carbinol nitrite in the vessel quantitatively. The efficient charges and coordinates of ions in a methyl phenyl carbinol nitrite molecule (left-handed or right-handed) are listed in the table I [16]. The unit of $R$ is angstrom and the charges are given in electrostatic unit.

TABLE I. The charge and spherical coordinates of atoms in methyl phenyl carbinol nitrite.

| atom   | $H_0$   | $H_1$  | $H_2$  | $H_3$  | $C_1$ | $C_2$  |
|--------|---------|--------|--------|--------|-------|--------|
| charge | 0.28    | 0.28   | 0.28   | 0.28   | -0.28 | -0.84  |
| $R$    | 1.13256 | 2.2104 | 2.2104 | 2.1967 | 0.00  | 0.49   |
| $\Theta$ | 1.9036 | 2.0969 | 2.0969 | 1.3877 | 0.00  | 3.1416 |
| $\Phi$ | 5.6589  | 2.0969 | 1.0246 | 1.5708 | 0.00  | 0.00   |

Through a complex computation of the central field approximation, we get the sum of the products of matrix elements for valence electrons in a methyl phenyl carbinol nitrite molecule

$$\sum_{i=1}^{N_3}\sum_{n\neq 0}\frac{\mathrm{Re}\{\langle\psi_{ni}|\mathbf{r}_i|\psi_{0i}\rangle\langle\psi_{0i}|\xi(\mathbf{r}_i)\mathbf{l}_i|\psi_{ni}\rangle\}}{E_{0i}-E_{ni}} \approx 4\times 10^{-18}. \quad (11)$$

In equation (11), Re{ } denotes the retention of the real part of the products of matrix elements, $E_{0i}-E_{ni}\approx 1$ eV and $N_3=48$ is taken for a methyl phenyl carbinol nitrite molecule. Let the dimensions of the copper cube and the cubical vessel be $0.08\times 0.08\times 0.01\mathrm{m}^3$ and $0.08\times 0.08\times 0.002\mathrm{m}^3$ respectively, the inter-plane distance between the copper cube and the vessel be 0.5 millimeter. The nucleon number density of the copper cube $\rho_n$ is about $5.4\times10^{30}$ m$^{-3}$ and the molecule number density in the vessel $\rho_M$ is about $4.427\times10^{27}$ m$^{-3}$. With a simple integration over the vessel and the copper cube, we get the magnitude of the force to be about $-2.36\times10^{25} g_s g_p$ for $\lambda = 10^{-3}$ m. It's determined by the magnitudes of scalar and pseudo-scalar coupling constants. The up-to-date measured result is $g_s g_p /(\hbar c) < 1.5\times10^{-24}$ for $\lambda = 10^{-3} m$ [3]. So the magnitude of the force between such copper cube and vessel of chiral molecules is less than $1.12\times10^{-24}$ N. Theoretically, this force may violate the test of the equivalence principle using a torsion pendulum in which the test masses are two vessels of homochiral molecules with opposite chirality.

**IV. CONCLUSIONS**

We demonstrated that the interaction mediated by exchanging light particles, such as axions, cooperates with the spin-orbit interaction to produce a parity-violating macroscopic force between chiral molecules and achiral source mass. Like optical rotation, this force is equal in magnitude but opposite in sign for molecules with the different chirality.

In order to measure this force, we calculated the magnitude of the force between a copper cube and homochiral molecules (methyl phenyl carbinol nitrite) in a cubical vessel using the central field approximation. It is shown that this effect should conceivably be detectable in experiments and it may be a possible cause of some chiral-dependent phenomena and the violation of the equivalence principle.

In our calculation, the dimensions of the copper cube and the cubical vessel and the inter-plane distance between them are so chosen as to fit the needs of laboratory experiment. The nucleon number density of the copper cube and the molecule number density in the vessel are estimated applicably. For the interaction range $\lambda = 10^{-3}$ m, the force is estimated to be $-2.36\times10^{25} g_s g_p$. According to the limits of the scalar and pseudo-scalar coupling constants published in previous work [3], the force is less than $1.12\times10^{-24}$ N, which is still far beyond the up-to-date detection limit in laboratory [17].

However, the spin-dependent force encountered for different chiral molecules may vary by a few order-of-magnitudes. Moreover, there may exist some kinds of chiral molecules for which the interaction are not so small. Due to the central field approximation used in calculation, the force may be underestimated because of the single-center theorem [12]. Thus the actual magnitude of the force may be greater than that we obtained. In order to obtain a more convincing result, we will improve our calculation with more applicable methods and try to further confirm its impact on the experimental testing of the equivalence principle.

**ACKNOWLEDGMENTS**

We would like to thank Dr. Chengang Shao for helpful discussions. This study was supported by National Basic Research Program of China (Grant No. 2003CB716300) and the National Natural Science Foundation of China (No. 10121503).

[1] R. H. Dicke, Rev. Mod. Phys. **29**, 355(1957); C. M. Will, Phys. Rep. **113**, 345 (1984); T. W. Darling, F. Rossi, G. I. Opat, and G. F. Moorhead, Rev. Mod. Phys. **64**, 237(1992); T. Damour, Classical Quantum Gravity **13**, A33(1996); J. Luo, Y. X. Nie, Y. Z. Zhang, and Z.B. Zhou, Phys. Rev. D **5**, 042005(2002); T. Damour, F. Piazza, and G. Veneziano, Phys. Rev. Lett. **89**, 081601(2002); Z. B. Zhou, J. Luo, Y. Z. Zhang, et.al., Rev. D **66**, 022002(2002); S. L. Dubovsky and V. A. Rubakov, Phys. Rev. D **67**, 104014(2003); Luca Amendola and Claudia Quercellini, Phys. Rev. Lett. **92**, 181102(2004); A. Füzfa and J. -M. Alimi, Phys. Rev. Lett. **97**, 061301(2006); S. Schlamminger, K. -Y. Choi and E. G. Adelberger. Phys. Rev. Lett. **100**, 041101(2008).

[2] M. Pospelov, Phys. Rev. D **58**, 097703(1998) ; M. Giovannini, Phys. Rev. D **59**, 063503(1999); Y. Itin and F. W. Hehl, Phys. Rev. D **68**, 127701(2003);S. M. Barr and B. Kyae, Phys. Rev. D **71**, 055006(2005).

[3] A. N. Youdin, D. Krause, and L. R. Hunter et al., Phys. Rev. Lett. **77**, 2170 (1996); W. T. Ni, S. S. Pan, H. C. Yeh, L. S. Hou, and J. Wan, Phys. Rev. Lett. **82**, 2439(1999); G. D. Hammond, C. C. Speake, C. Trenkel et, al., Phys. Rev. Lett. **98**, 081101(2007).

[4] J. E. Moody and F. Wilczek, Phys. Rev. D **30**, 130(1984).

[5] Biot, Bull soc.philomath. 190(1815).

[6] F. C. Michel, Phys. Rev. **138**, B408(1965); V. S. Letokhov, Phys. Lett. **53** A275(1975); D.K.Kondepudi and G. W. Nelson, Nature.**314**, 438(1985); A. Bakasov,T.-K. Ha, and M. Quack, J. Chem. Phys. **109**, 7263(1998); R. Zanasi et al., Phys. Rev. E **59**, 3382(1999); J. K. Laerdahl and P. Schwerdtfeger, Phys. Rev. A **60**, 4439(1999).

[7] H. Eyring, J. Walter, G. E. Kimball, Quantum Chemistry, Wiley, New York, 1944.

[8] J. C. Slater, Quantum Theory of Matter, 2nd ed. (McGraw Hill, NewYork, 1968); I. I. Sobelman, Introduction to the Theory of Atomic Spectra (Pergamon, Oxford, 972); B. Edlen, Atomic Spectra, Encyclopedia of Physics, (Springer-Verlag, Berlin,1964), Vol.XXVII, p. 80.

[9] Landau and Lifshitz, Quantum mechanics, 2nd ed. Ch.X(Pergamon, London, 1965).

[10] M. A. Bouchiat and C. Bouchiat, J. Phys. (Paris) **35**, 899(1974), **36**, 493(1975); B. Y. Zel'dovich et al., Zh. Eksp. Teor. Fiz. Pis' ma Red. **25**, 106(1977).

[11] S. F. Mason and G. E. Tranter, Molecular Physics.**53**, 1091(1984).

[12] R. A. Hegstrom, D. W. Rein, P. G. H. Sardars, J. Chem. Phys.**73**, 2329(1980).

[13] R. A. Harris and L. Stodosky, Phys. Lett. **78B**, 313(1978).

[14] G. L. Tan, M. F. Lemon, D. J. Jones, and R. H. French , Phys. Rev. B **72**, 205117(2005).
[15] P. Rudra, J. Math. Phys. **6**, 1278(1965).
[16] E. U. Condon, Mod. Phys. Rev. **9**, 432(1937).
[17] Liang-Cheng Tu, Sheng-Guo Guan, and Jun Luo, et,al. Phys. Rev. Lett. **98**, 201101(2007）.